\documentclass[twocolumn,english]{article}
\usepackage[T1]{fontenc}
\usepackage{color}
\usepackage{graphicx}
\usepackage{amssymb}

\makeatletter

\providecommand{\tabularnewline}{\\}
\usepackage{a4}
\input{epsfig.sty}
\def\fnum@table{\tablename~{\bf\thetable}}
\def\fnum@figure{\figurename~{\bf\thefigure}}
\def\tablename{\footnotesize{\bf Table}}
\def\figurename{\footnotesize{\bf Figure}}

\def\be{\begin{equation}}
\def\ee{\end{equation}}

\makeatother

\usepackage{babel}

\begin{document}

\title{\textbf{LHC data on inelastic diffraction
and   uncertainties in the predictions
for longitudinal EAS development}}

\author{\textbf{S. Ostapchenko}%
\thanks{e-mail: sergei@tf.phys.ntnu.no%
} \\
\textit{\small D.V. Skobeltsyn Institute of Nuclear Physics, Moscow
State University, 119991 Moscow, Russia}\textit{ }\\
}

\maketitle
\begin{center}
\textbf{\large Abstract}
\par\end{center}{\large \par}

Present status of experimental studies of inelastic diffraction
at the Large Hadron Collider is analysed. Impact of the current
uncertainties concerning the diffraction rate on the predicted
extensive air shower development is investigated. Relation to
studies of the primary composition of ultrahigh energy cosmic
rays is illustrated by comparing numerical simulation results
to the data of the Telescope Array experiment on the distributions
of the shower maximum position.

\section{{\normalsize Introduction}\label{intro.sec}}

Among the most outstanding problems in the high energy cosmic ray
(CR) field is the determination of the composition of ultra-high energy
cosmic rays (UHECR). The corresponding experimental studies are
based on the so-called extensive air shower (EAS) technique \cite{nagano}:
the properties of primary cosmic ray particles (protons and nuclei)
are reconstructed from measured characteristics of nuclear-electromagnetic
cascades induced by them in the atmosphere. Respectively,
the obtained results depend strongly on the correctness of the Monte
Carlo (MC) procedures used for numerical simulations of air showers,
notably, on the models of hadronic interactions, employed in such
simulation programs. This brings an additional source of uncertainty
for the experimental results - as such models are largely phenomenological
ones and have to be extrapolated from accelerator energies, where
they are calibrated, to the much higher UHECR energies \cite{ehp11}.

In this respect, experimental data on proton-proton interactions,
obtained at the Large Hadron Collider (LHC) at the highest thus far
collision energies, prove to be invaluable for improving EAS simulation
procedures, reducing thereby the above-discussed uncertainty in CR
studies. Importantly, a comparison of the predictions of hadronic
MC generators with LHC data revealed that hadronic interaction models
used in the CR field provide adequate enough description of the main
features of proton-proton interactions \cite{david11}. Moreover, a
number of model updates emerged recently \cite{ost11,epos-lhc,sibyll2.2},
which included new fine tunes of model parameters, based on LHC data,
as well as improvements in the underlying theoretical framework.
Nevertheless, there remain considerable differences between the model
predictions for basic EAS characteristics, which constitute a serious
obstacle for precise studies of the UHECR composition \cite{pierog13a}.

Presently, mass composition of high energy cosmic rays is studied
by two different techniques \cite{nagano}: i) via measurements
of lateral densities of all charged particles and of muons only by
ground-based detectors; ii) via studies of the longitudinal shower
development with fluorescence detectors. While both methods can generally
be powerful enough for determining the UHECR composition, the progress
of the ground-based studies is presently hampered by the strong contradiction
between the data of the Pierre Auger Observatory (PAO) for EAS muon
content at primary CR energies $E_{0}>3\cdot10^{18}$ eV and the respective
predictions of the shower simulation procedures \cite{pao-muons}. The reported
large (factor $1.3-1.6$) discrepancy between the PAO data and the simulation
results is especially surprising in view of the above-mentioned calibration
of hadronic interaction models to LHC data. Moreover, no such contradiction
has been observed by the KASCADE-Grande experiment at slightly lower
energies ($E_{0}<10^{18}$ eV) \cite{kaskade-muons} or by the Yakutsk experiment
in the UHECR energy range \cite{glushkov13}. In view of this confusing
situation, we shall restrict our analysis to the observables related to the
longitudinal EAS development, namely, to the position of
the shower maximum $X_{\max}$ (the depth in the atmosphere where
a maximal number of ionizing particles is observed) and its distribution.

Remarkably, the shower maximum position for proton-induced EAS depends
mostly on characteristics of the interaction of the primary cosmic
ray particle with air nuclei, notably on the respective inelastic
cross section $\sigma_{p{\rm -air}}^{{\rm inel}}$. Hence, recent
precise measurements (with per cent level accuracy) of the total, elastic,
and inelastic proton-proton cross sections at $\sqrt s=7$ and 8 TeV
by the TOTEM experiment 
\cite{totem11,totem13,totem13a,totem13b,totem13c}
provide extremely important constrains for the respective model predictions
- as $\sigma_{p{\rm -air}}^{{\rm inel}}$ can thus be calculated in
the framework of the Glauber-Gribov approach \cite{glauber58,gribov68}. 

Unfortunately, additional uncertainties arise from the treatment of
inelastic diffraction in hadronic interaction models, which impacts
model predictions for $X_{\max}$ in two ways. First, inelastic diffraction
is intimately related to the inelastic screening effect for the calculated
cross sections of hadron-proton and hadron-nucleus interactions \cite{gribov69}:
a higher rate of diffraction dissociation is accompanied
by stronger screening effects which give rise to a smaller
hadron-nucleus cross section predicted (e.g.~\cite{parsons11}). Secondly,
the rate of inelastic diffraction largely dominates model predictions
for the so-called inelasticity $K_{p{\rm -air}}^{{\rm inel}}$ - the
relative energy loss of the leading (most energetic) secondary nucleon
in $p-$air collisions. For example, in the target diffraction process
at very high energies the leading proton loses only a tiny fraction
of its energy: $\Delta E/E_{0}\simeq\exp(-\Delta y)\ll1$, where $\Delta y$
is the size of the rapidity gap 
 between the struck proton and the most energetic
secondary hadron produced in the diffractive excitation of the target
nucleus.\footnote{Typically, $\Delta y \gtrsim\ln \sqrt{s}$.}
 Thus, enhancing target diffraction is equivalent to effectively
reducing the total inelastic cross section $\sigma_{p{\rm -air}}^{{\rm inel}}$.
As both above-discussed effects work in the same direction, one has
a simple ``rule-of-thumb'': higher diffraction rate corresponds to
a slower EAS development (deeper shower maximum) and vise verse.

In the following, we are going to investigate the impact of the
present experimental uncertainties concerning the rate of the inelastic 
diffraction in hadronic collisions on model predictions for the longitudinal 
EAS development and on the related  studies of the UHECR composition.
Our analysis will be based on the QGSJET-II-04 model \cite{ost11}
 which is characterized by a microscopic treatment
of nonlinear interaction effects in hadronic collisions and has thus
a much higher predictive power for various interaction characteristics,
notably for diffractive cross sections, compared to other MC generators.

The paper is organized as follows. In Section \ref{lhc-diffr.sec}, we review
recent LHC results on inelastic diffraction and illustrate certain tensions
between the data of different experiments by comparing them to predictions of
the QGSJET-II-04 model. Additionally, we perform two additional tunes of the
model parameters, designed to fit better different sets of measurements.
The predictions of these two alternative model versions for the average
$X_{\max}$ and its fluctuations are compared in  Section \ref {xmax.sec}
and the respective differences are regarded as the corresponding model
uncertainty. Further, we illustrate the potential impact of this uncertainty 
on UHECR composition studies    by applying the alternative
model versions to fitting $X_{\max}$ distributions measured by the 
Telescope Array experiment. We conclude in Section \ref{summary.sec}.

\section{{\normalsize LHC results on inelastic diffraction}\label{lhc-diffr.sec}}

Studies of the inelastic diffraction constitute an important part
of the experimental program at the Large Hadron Collider, with important
results obtained by the ALICE \cite{alice-difr}, ATLAS \cite{atlas-lrg}, 
CMS \cite{cms-difr}, and TOTEM
\cite{totem13a,totem-dd,totem-sd} experiments.
Unfortunately, at present stage there exist certain tensions
between TOTEM measurements of diffractive cross sections and the respective
CMS and ATLAS results, as already discussed in Ref.\ \cite{khoze13}.

\begin{figure*}[tbh]
\includegraphics[width=1\textwidth]{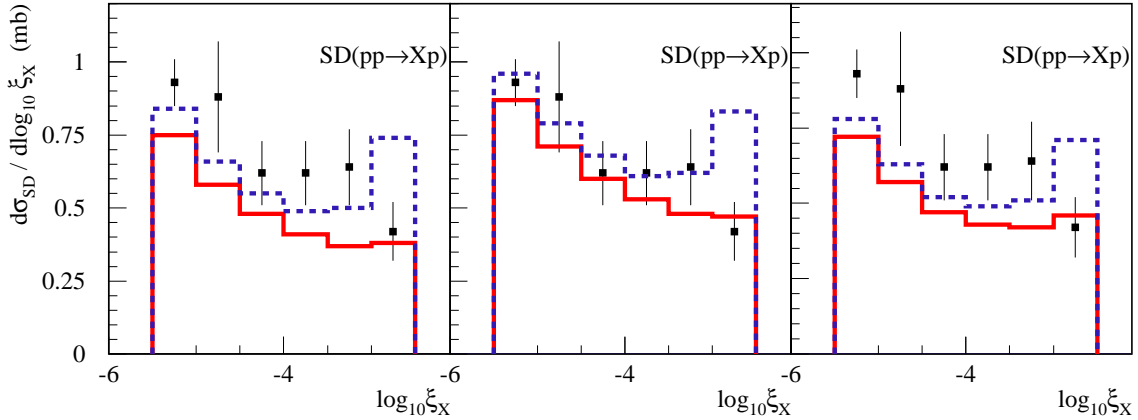}
\caption{Calculated $\xi_{X}\equiv M_{X}^{2}/s$ dependence of $\sigma_{pp\rightarrow Xp}^{{\rm SD}}$
at $\sqrt{s}=7$ TeV (full histograms) compared to CMS data \cite{cms-difr}
(points) for QGSJET-II-04 (left), option SD+ (middle), and option
SD- (right). Same dependence for $\sigma_{pp\rightarrow Xp}^{{\rm SD}}+\sigma_{pp\rightarrow XY}^{{\rm DD}}$
($M_{Y}<3$ GeV) is shown by dashed histograms.\label{fig:SD-cms}}
\end{figure*}

To get a deeper insight into the problem, 
we start by comparing the results of the TOTEM and CMS experiments for
the cross section of single diffraction $\sigma_{pp}^{{\rm SD}}$,
for different ranges of mass $M_{X}$ of diffractive states produced,
with the predictions of the QGSJET-II-04 model: 
Tables  \ref{tab: SD-totem}, \ref{tab: SD-cms}
\begin{table*}[t]
\begin{tabular*}{1\textwidth}{@{\extracolsep{\fill}}lccccc}
\hline 
$M_{X}$ range & $<3.4$ GeV & $3.4-1100$ GeV & $3.4-7$ GeV & $7-350$ GeV &
 $350-1100$ GeV\tabularnewline
\hline 
\hline 
TOTEM \cite{totem13a,totem-sd} & $2.62\pm2.17$ & $6.5\pm1.3$ & $\simeq 1.8$
 & $\simeq 3.3$ & $\simeq 1.4$\tabularnewline
QGSJET-II-04 & 3.9 & 7.2 & 1.9 & 3.9 & 1.5\tabularnewline
option SD+ & 3.2 & 8.2 & 1.8 & 4.7 & 1.7\tabularnewline
option SD- & 2.6 & 7.2 & 1.6 & 3.9 & 1.7\tabularnewline
\hline 
\end{tabular*}\caption{$\sigma_{pp}^{{\rm SD}}$ (mb) at $\sqrt{s}=7$ TeV for different
ranges of mass $M_{X}$ of diffractive states produced.\label{tab: SD-totem}}
\end{table*}
{\large }
\begin{table}[tbh]
{\small }%
\begin{tabular*}{0.49\textwidth}{@{\extracolsep{\fill}}cccc}
\hline 
{\scriptsize CMS \cite{cms-difr}} & {\scriptsize QGSJET-II-04} & {\scriptsize option SD+} & {\scriptsize option SD-}\tabularnewline
\hline 
\hline 
$4.3\pm0.6$ & 3.0 & 3.7 & 3.1\tabularnewline
\hline 
\end{tabular*}{\small \caption{$\sigma_{pp}^{{\rm SD}}$ (mb) at $\sqrt{s}=7$ TeV for $12<M_{X}<394$
GeV.\label{tab: SD-cms}}
}
\end{table}
 and Fig.~\ref{fig:SD-cms} (left panel).
  It is easy to see that the $M_{X}$-dependencies
observed by the two experiments qualitatively agree with each other 
and with the
model predictions. However, the absolute rates of the inelastic diffraction
measured by TOTEM and CMS are noticeably different: while the results
of QGSJET-II-04 agree with TOTEM values within the reported experimental
uncertainties, the model predictions appear to be in variance with
the CMS measurements, lacking some $30$\% of $\sigma_{pp}^{{\rm SD}}$
observed by CMS. The discussed contradiction is surprising taking
the fact that the kinematic range studied by CMS ($12\:{\rm GeV}<M_{X}<394\:{\rm GeV}$)
is fully covered by TOTEM ($3.4<M_{X}<1100$ GeV). In principle, as the
CMS analysis is based on the rapidity gap technique, its results depend
noticeably on model-dependent corrections. A relevant example is the 
subtraction of the contribution of double diffraction, 
when one of the diffractively
excited states is characterized by a small   mass ($M_{Y}<3$
GeV) and thus remains unobserved by the experimental apparatus. Such a
contribution is potentially dangerous as the MC generators used in
the analysis lack any specific treatment for low mass diffraction,
notably for the diffractive production of low mass resonance states
(e.g.~$N^{*}$), as stressed previously in \cite{ost11b}. As an illustration,
we plot in Fig.~\ref{fig:SD-cms} (left panel)
 the results for the sum of $\sigma_{pp\rightarrow Xp}^{{\rm SD}}$
and $\sigma_{pp\rightarrow XY}^{{\rm DD}}(M_{Y}<3\:{\rm GeV})$ for
the $M_{X}$-range studied by CMS (blue dashed line). 
However, even in that case one
is unable to reach a satisfactory agreement between the model results
and the CMS data.

\begin{figure}[tbh]
\includegraphics[width=0.49\textwidth]{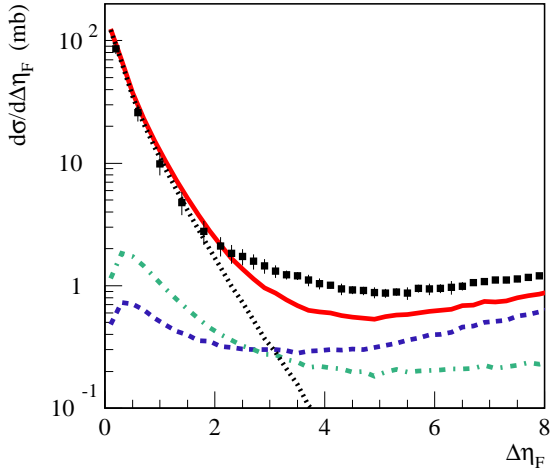}
\caption{Cross section for forward rapidity gap production in $pp$-collisions
at $\sqrt{s}=7$ TeV, calculated with QGSJET-II-04   (solid
line), in comparison with ATLAS data \cite{atlas-lrg} (points).
 Separately shown
contributions from single diffractive (dashed), double diffractive
(dot-dashed), and nondiffractive (dotted) interactions.\label{fig:rapgap}}
\end{figure}

Moreover, comparing in Fig.~\ref{fig:rapgap}
 the model predictions for the cross section for forward rapidity
gap production $d\sigma_{pp}/d\Delta\eta_{{\rm F}}$, $\Delta\eta_{{\rm F}}$
being the forward rapidity gap size, with respective ATLAS data \cite{atlas-lrg},
we see the same level of disagreement ($30-40$\%) despite the fact
that both single and double diffraction processes contribute to the
studied cross sections. A potential way out of the contradiction is
to assume that QGSJET-II-04 seriously underestimates the contribution
of double diffraction and that the latter dominates  $d\sigma_{pp}/d\Delta\eta_{{\rm F}}$
and also contaminates noticeably $\sigma_{pp}^{{\rm SD}}$ measured
by CMS. Comparing the prediction of the model for the cross section
of high mass diffraction $\sigma_{pp\rightarrow XY}^{{\rm DD}}$ ($M_{X},M_{Y}>10$
GeV), with the rapidity gap between the two diffractive states $\Delta\eta>3$,
with the CMS data in Fig.~\ref{tab: DD-cms}
\begin{figure}[tbh]
\includegraphics[width=0.49\textwidth]{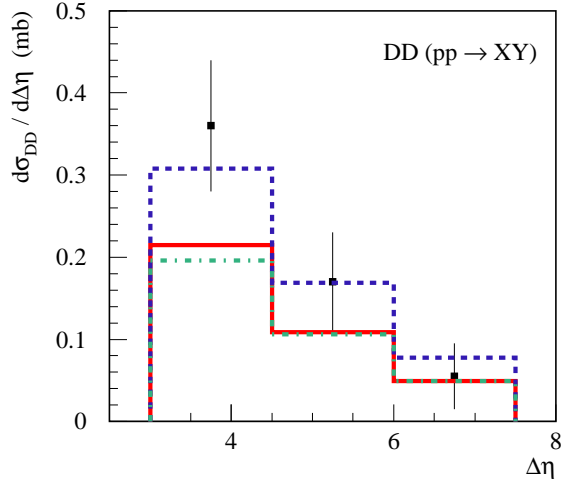}
\caption{Calculated $\sigma_{pp\rightarrow XY}^{{\rm DD}}$ (mb) at $\sqrt{s}=7$
TeV as a function of the rapidity gap size $\Delta\eta=-\log(M_{X}^{2}M_{Y}^{2}/(s\cdot m_{p}^{2}))$
($m_{p}$ being the proton mass) for $M_{X},M_{Y}>10$ GeV compared
to CMS data \cite{cms-difr} (points) for QGSJET-II-04 (solid), option
SD+ (dashed), and option SD- (dot-dashed).\label{fig:DD}}

\end{figure}
 and Table \ref{tab: DD-cms},{\large }
\begin{table}[tbh]
{\small }%
\begin{tabular*}{0.49\textwidth}{@{\extracolsep{\fill}}cccc}
\hline 
{\scriptsize CMS \cite{cms-difr}} & {\scriptsize QGSJET-II-04} & {\scriptsize option SD+} & {\scriptsize option SD-}\tabularnewline
\hline 
\hline 
{\scriptsize $0.93\pm0.01_{-0.22}^{+0.26}\!\!\!\!\!\!\!\!\!$} & {\scriptsize 0.57} & {\scriptsize 0.85} & {\scriptsize 0.54}\tabularnewline
\hline 
\end{tabular*}\caption{$\sigma_{pp\rightarrow XY}^{{\rm DD}}$ (mb) at $\sqrt{s}=7$ TeV
for $M_{X},M_{Y}>10$ GeV and $\Delta\eta>3$.\label{tab: DD-cms}}
\end{table}
 we find indeed a rather large ($\sim40$\%) disagreement which is,
however, insufficient to explain the above-discussed discrepancies
(c.f.~the contribution of double diffraction to $d\sigma_{pp}/d\Delta\eta_{{\rm F}}$
in Fig.~\ref{fig:rapgap}). Moreover, the model prediction for the
rate of double diffraction proves to be in a good agreement with TOTEM
measurements, see Table \ref{tab: DD-totem}.{\large }
\begin{table}[tbh]
{\small }%
\begin{tabular*}{0.49\textwidth}{@{\extracolsep{\fill}}cccc}
\hline 
{\scriptsize TOTEM \cite{totem-dd}} & {\scriptsize QGSJET-II-04} & {\scriptsize option SD+} & {\scriptsize option SD-}\tabularnewline
\hline 
\hline 
{\scriptsize $116\pm25$} & {\scriptsize 134} & {\scriptsize 152} & {\scriptsize 102}\tabularnewline
\hline 
\end{tabular*}\caption{$\sigma_{pp}^{{\rm DD}}$ ($\mu$b) at $\sqrt{s}=7$ TeV for the minimum
pseudorapidity of produced hadrons $4.7<|\eta_{\min}|<6.5$.\label{tab: DD-totem}}
\end{table}
{\large{} }{\large \par}

\begin{figure*}[t]
\includegraphics[width=1\textwidth]{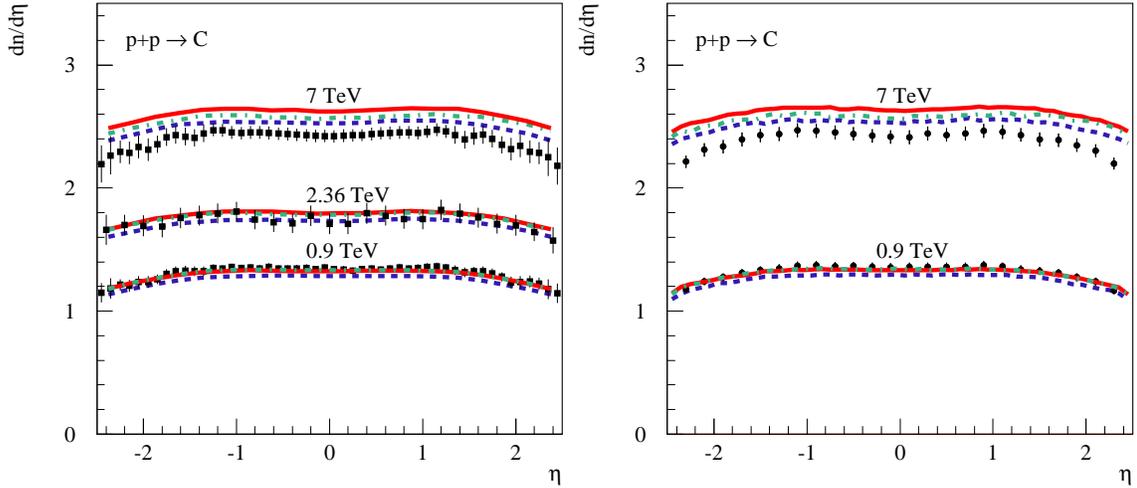}
\caption{Pseudorapidity density of charged hadrons of transverse momentum $p_{t}>0.5$
GeV produced in $pp-$collisions at $\sqrt{s}=0.9$, 2.36, and 7 TeV
(as indicated in the plots) as calculated using the default QGSJET-II-04
model (solid), option SD+ (dashed), and option SD- (dot-dashed) compared
to experimental data from ATLAS \cite{atlas11} (squares) and CMS \cite{cms11}
(circles).\label{fig:eta}}
\end{figure*}

Generally, the TOTEM experiment has a good potential for reliable
measurements of diffractive cross sections - thanks to the Roman Pot
technique employed. However, at the present stage we have no choice
but to regard the differences between the preliminary TOTEM results
and the ones of the CMS and ATLAS experiments as the experimental
uncertainty for the diffraction rates. In the next Section, we are
going to investigate the impact of this uncertainty on the model predictions
for $X_{\max}$ and for related studies of UHECR composition. To this
end, we create two additional versions of the model, with alternative
tunes of its parameters.
\begin{figure}[t]
\includegraphics[width=0.49\textwidth]{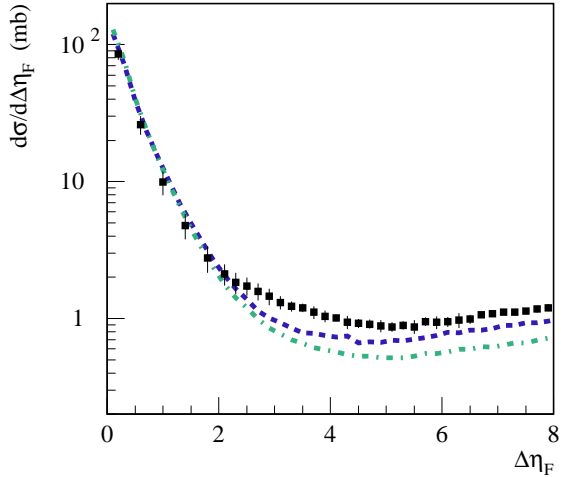}
\caption{Cross section for forward rapidity gap production in $pp$-collisions
at $\sqrt{s}=7$ TeV, calculated with the options SD+  (dashed) and SD-
(dot-dashed), compared to ATLAS data
 \cite{atlas-lrg} (points).\label{fig:rapgap-optns}}
\end{figure}
\begin{figure}[t]
\includegraphics[width=0.49\textwidth]{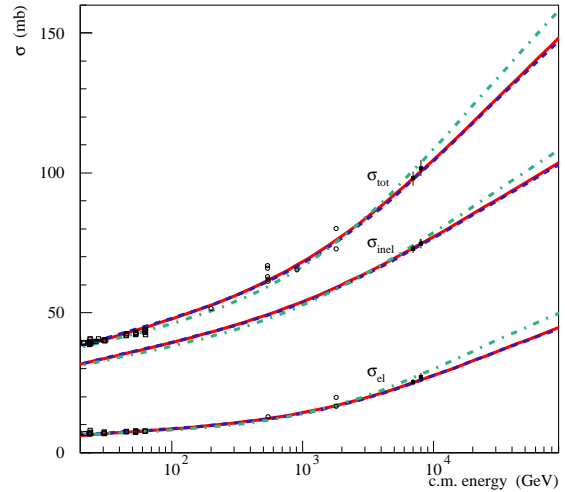}
\caption{Energy dependence of total, inelastic, and elastic $pp$ cross sections
as calculated using the default QGSJET-II-04 model (solid), option
SD+ (dashed), and option SD- (dot-dashed). Experimental data are from
\cite{totem13b,totem13c,pdg}.\label{fig:sigmas}}

\end{figure}
 In one case, referred below as ``option
SD+'', we enhance the contribution of high mass 
diffraction\footnote{Technically,  a higher
rate for high mass diffraction is obtained by increasing the value of the
triple-Pomeron coupling in the model, which thus impacts both single and
double diffraction processes \cite{ost11,ost10}. In turn, the rate of low mass 
diffraction is governed by the structure of Good-Walker diffractive
eigenstates, notably, by their relative interaction strengths 
\cite{parsons11}.} in order
to reach a reasonable agreement with ATLAS and CMS --
 see  Figs.~\ref{fig:SD-cms} (middle   panel),
 \ref{fig:DD}, and \ref{fig:rapgap-optns};
also Tables \ref{tab: SD-cms} and \ref{tab: DD-cms}. At the same time, we  
slightly reduce the rate of low mass diffraction -- in order to
soften the obtained disagreement with TOTEM (Tables \ref{tab: SD-totem}
and \ref{tab: DD-totem}). Alternatively, we choose to fit more closely
the TOTEM result for the  low mass diffraction cross section \cite{totem13a}
by seriously reducing the respective contribution  (by as much as 30\%), 
 while keeping more or less the same rate for
high mass diffraction (``option SD-''). The respective results are compared
to the   TOTEM, CMS, and ATLAS data in Tables \ref{tab: SD-totem}-\ref{tab: DD-totem} and  Figs.~\ref{fig:SD-cms} (right   panel),
 \ref{fig:DD}, \ref{fig:rapgap-optns}.
 In addition, the option SD+
is characterized by a slightly slower energy rise of the total and
inelastic cross sections, while the opposite is true for the option
SD-, both within the experimental uncertainties -- Fig.~\ref{fig:sigmas}.
 For both versions, model parameters are tuned in such a way that
particle production in the central rapidity range remains similar
to the original QGSJET-II-04, as illustrated in Fig.~\ref{fig:eta}.
\begin{figure*}[t]
\includegraphics[width=1\textwidth]{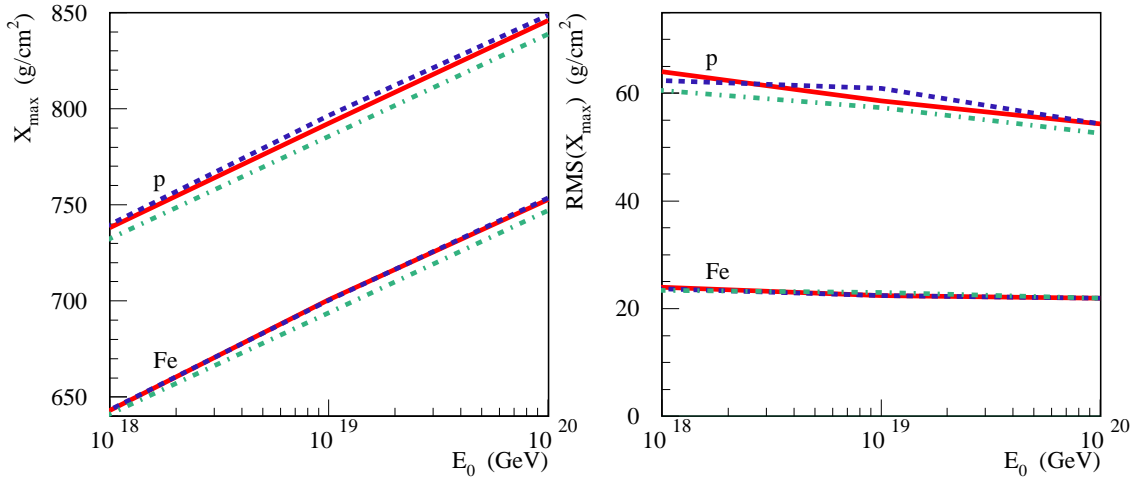}
\caption{Average $X_{\max}$ (left) and RMS($X_{\max}$) (right) for the default
QGSJET-II-04 model (solid), option SD+ (dashed), and option SD- (dot-dashed).\label{fig:xmax-rms-options}}
\end{figure*}

\section{{\normalsize Impact on $X_{\max}$ predictions and on UHECR composition
studies}\label{xmax.sec}}

Now we apply both the original QGSJET-II-04 and the two alternative model 
versions to air shower simulations, using the CONEX program \cite{conex}.
The obtained primary energy dependencies of the predicted average $X_{\max}$
and of the corresponding shower maximum distribution width RMS($X_{\max}$) 
for the three models considered are presented in Fig.~\ref{fig:xmax-rms-options}.
 The plots demonstrate how the present experimental uncertainties
 concerning the rate of inelastic diffraction 
 project themselves on the predicted EAS characteristics.
 While the respective uncertainties for RMS($X_{\max}$)   prove to be
 negligibly small (less than  3 g/cm$^{2}$), those for the average shower
 maximum position appear to be quite sizable:   $X_{\max}$ predictions for
 the two alternative model versions (options SD+ and SD-) differ from each
 other by some  10 g/cm$^{2}$. While being already smaller than typical
 experimental inaccuracies of  $X_{\max}$ measurements ($15-20\;{\rm g/cm}^2$),
 these model uncertainties may noticeably degrade the accuracy of UHECR
 composition studies.
 
To illustrate the latter point, we apply the above-described model versions
to a simplified analysis of the 
cosmic ray composition in the very high energy range, using the data of the
Telescope Array (TA) experiment \cite{ta-xmax}. 
In principle, as demonstrated already in
Ref.\ \cite{aloisio08}, the width of $X_{\max}$-distributions
RMS($X_{\max}$) could be a very convenient tool for  CR composition
 studies: the quantity is practically independent on any other details of 
 interaction models used for EAS simulations except the predicted total 
 inelastic cross section and the inelastic diffraction rate. 
 However, experimental determination of RMS($X_{\max}$) is somewhat 
 challenging due to its sensitivity to data quality cuts employed in a 
 particular analysis and to other details of experimental procedures.
 Therefore, correcting for such effects, inherent for a particular experiment,
 is a nontrivial problem. Hence, we apply here a more standard method,
 trying to deduce the primary composition from fitting the measured
 $X_{\max}$ distributions by   simulated ones, for different mixtures
 of primary CR particles.
\begin{figure*}[t]
\includegraphics[width=1\textwidth]{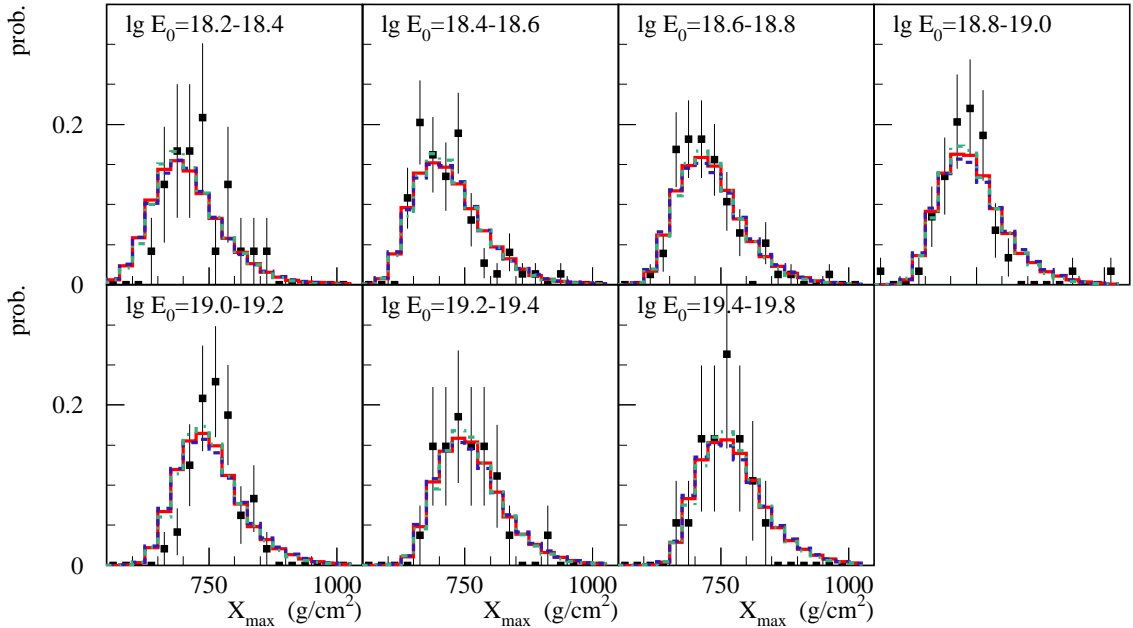}
\caption{$X_{\max}$ distributions
 measured by the Telescope Array experiment \cite{ta-xmax}
 compared to calculations with the default QGSJET-II-04
model (solid), option SD+ (dashed), and option SD- (dot-dashed) for
the fitted primary compositions from Table \ref{tab:fit-TA}. 
\label{fig:fit-TA-distr}}
\end{figure*}%
 
As the measured $X_{\max}$ distributions  are  influenced by experimental 
measurement and reconstruction procedures, 
the consistency requires  the output
of EAS simulation procedures   to be processed through the same analysis
and reconstruction chains as the respective experimental data.
In this work, we choose an alternative way: we mimic the above-discussed
effects by applying a systematic shift $\Delta X_{max}$ and an additional
Gaussian smearing $\Delta\sigma$ to $X_{\max}$ distributions obtained
from EAS simulations, as described in more detail in the Appendix.
Using this method, we  fit $X_{\max}$ distributions
measured by the Telescope Array experiment in a number of
primary energy intervals, using a two-component mixture ($p$
plus $Fe$) for the primary CR composition and assuming the relative
abundances $d_{i}$ ($i=p,\, Fe$) to depend logarithmically on the
energy of the primary particle $E_{0}$:
\begin{eqnarray}
d_{p}(E_{0}) & = & d_{p}(1)+\left[d_{p}(100)-d_{p}(1)\right]\nonumber \\
 & \times & \lg(E_{0}/1\,{\rm EeV})/2\label{eq:comp-fit}\\
d_{Fe}(E_{0}) & = & 1-d_{p}(E_{0})\,.\nonumber 
\end{eqnarray}
Here $d_{p}(1)$ and $d_{p}(100)$ refer to proton abundances at 1
and 100 EeV respectively.

 The fitted primary abundances are presented in Table  \ref{tab:fit-TA},
\begin{table}[tbh]
\begin{tabular*}{0.49\textwidth}{@{\extracolsep{\fill}}cccc}
\hline 
 & $d_{p}(1)$ & $d_{p}(100)$ & $\chi^{2}$/d.o.f.\tabularnewline
\hline 
\hline 
QGSJET-II-04 & 0.79 & 0.77 & 35.6/33\tabularnewline
option SD+ & 0.77 & 0.75 & 41.4/33\tabularnewline
option SD- & 0.84 & 0.85 & 31.8/33\tabularnewline
\hline 
\end{tabular*}\caption{Parameters for the composition fit {[}Eq.~(\ref{eq:comp-fit}){]}
based on Telescope Array $X_{\max}$ data.\label{tab:fit-TA}}
\end{table}
 while the corresponding $X_{\max}$ distributions are shown in 
 Fig.~\ref{fig:fit-TA-distr}
 in comparison to the experimental data. We attempted also to fit
the data with a three-component composition mixture, adding either
helium or carbon nuclei as the third primary group, but haven't obtained
a significant improvement of the quality of the fits.

\begin{figure}[t]
\includegraphics[width=0.49\textwidth]{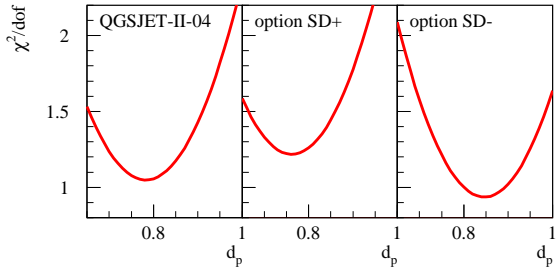}
\caption{Goodness of fits to the TA $X_{\max}$ distributions for an energy-independent
primary composition with proton fraction $d_{p}$, using different
interaction models (as indicated in the plots).\label{fig:chi2-TA}}
\end{figure}

It is easy to see that  the obtained fraction of primary iron nuclei is
 very sensitive
to the uncertainties studied in this work,   amounting to
 10\% difference between the options SD+ and SD-. 
One may equally well fit the data with an   energy-independent composition
mixture ($d_{p}(E_{0})=d_{p}={\rm const}$), as illustrated 
 in Fig.~~\ref{fig:chi2-TA}.
Here we see how the uncertainties related to inelastic diffraction rate 
may influence the interpretation of experimental data:
 while for the model option SD- the data are  
  consistent with  an almost pure proton composition in the energy
range $E_{0}=10^{18}-10^{20}$ eV, this is no longer valid for the option
SD+, in which case a substantial fraction of heavy nuclei is required.
These differences may have long-ranging consequences for astrophysical
interpretations of UHECR  data, e.g.\  for discriminating between
models for the transition from galactic to extragalactic cosmic ray
origin in the ultra-high energy range (see \cite{ber12} for a recent review).

\section{{\normalsize Summary}\label{summary.sec}}
We discussed recent LHC results on  inelastic
diffraction in $pp$ collisions  and demonstrated that there exists
a substantial uncertainty concerning the rate of diffractive collisions.
This latter projects itself on model-based calculations of 
the development of CR-induced extensive air showers in the atmosphere,
resulting in some 10 g/cm$^{2}$ uncertainty for the predicted shower maximum
position $X_{\max}$. Though being already smaller than the typical 
experimental precision for $X_{\max}$ measurements, this uncertainty may
noticeably degrade the accuracy of UHECR composition studies and,
as demonstrated above, can even seriously bias   astrophysical
 interpretations of cosmic ray data. Thus, further progress in 
 experimental studies of inelastic diffraction at the Large Hadron Collider
 is of utmost importance for the cosmic ray field.

One may question if there exist other uncertainties which impact model
predictions for the longitudinal air shower development. Unfortunately,
this is indeed the case: predicted  $X_{\max}$ depends noticeably on the
multiplicity of secondary particles in proton-air interactions \cite{ulrich10}.
 Present LHC data appear to be insufficient to fully remove this uncertainty
  due to a significant model dependence for the generalization from
  $pp$ to $pA$ collisions. In particular, it is the  smaller multiplicity
  of proton-nitrogen collisions, predicted by the EPOS-LHC model compared to
  QGSJET-II-04, which is the reason for deeper   $X_{\max}$ predicted by that
  model (by as much as 20 g/cm$^{2}$) \cite{pierog13a}.
   Thus, experimental studies of collisions
  of protons with light nuclei (nitrogen or oxygen) at LHC could be very
  useful for finally settling the issue.

\subsection*{{\normalsize Acknowledgments}}

The author is indebted to V.\ Berezinsky for motivating discussions which
stimulated this investigation. Useful discussions with M.\ Kacheriess,
 V.\ Khoze, T.\ Pierog, and M.\ Unger are gratefully acknowledged.

\section*{{\normalsize Appendix}}

Measured distributions of the shower maximum position $X_{\max}$
are influenced by experimental quality cuts and by reconstruction
procedures employed in experimental analysis. Therefore, to compare
numerical simulation results to the data and to perform an analysis
of the cosmic rays composition, the output of EAS simulation procedures
has to be processed through the same analysis and reconstruction chains
as the respective experimental data.

In this work, we choose an alternative way: we mimic the above-discussed
effects by applying a systematic shift $\Delta X_{max}$ and an additional
Gaussian smearing $\Delta\sigma$ to $X_{\max}$ distributions obtained
from EAS simulations. In case of TA data,
 the values of $\Delta X_{max}$
and $\Delta\sigma$ are defined via a least squares  minimization
of the difference between the so-modified $X_{\max}$ distributions
obtained with the QGSJET-II-03 model \cite{ost06} and the respective
simulation results of the Telescope Array collaboration,
 based on the same interaction model,
  which were obtained via processing the model predictions through
the complete experimental analysis and reconstruction chain \cite{ta-xmax}.
Subsequently, we apply the so-obtained shift and smearing parameters
to $X_{\max}$ distributions obtained both with the default QGSJET-II-04
model and with the two alternative model tunes described in the text
in order to compare the model results to experimental data. As all
the above-discussed models have more or less the same physics content
and their predicted $X_{\max}$ distributions have similar shapes,
we believe the procedure is accurate enough for the purposes of the
present investigation.
\begin{figure*}[t]
\includegraphics[width=1\textwidth,height=13.5cm]{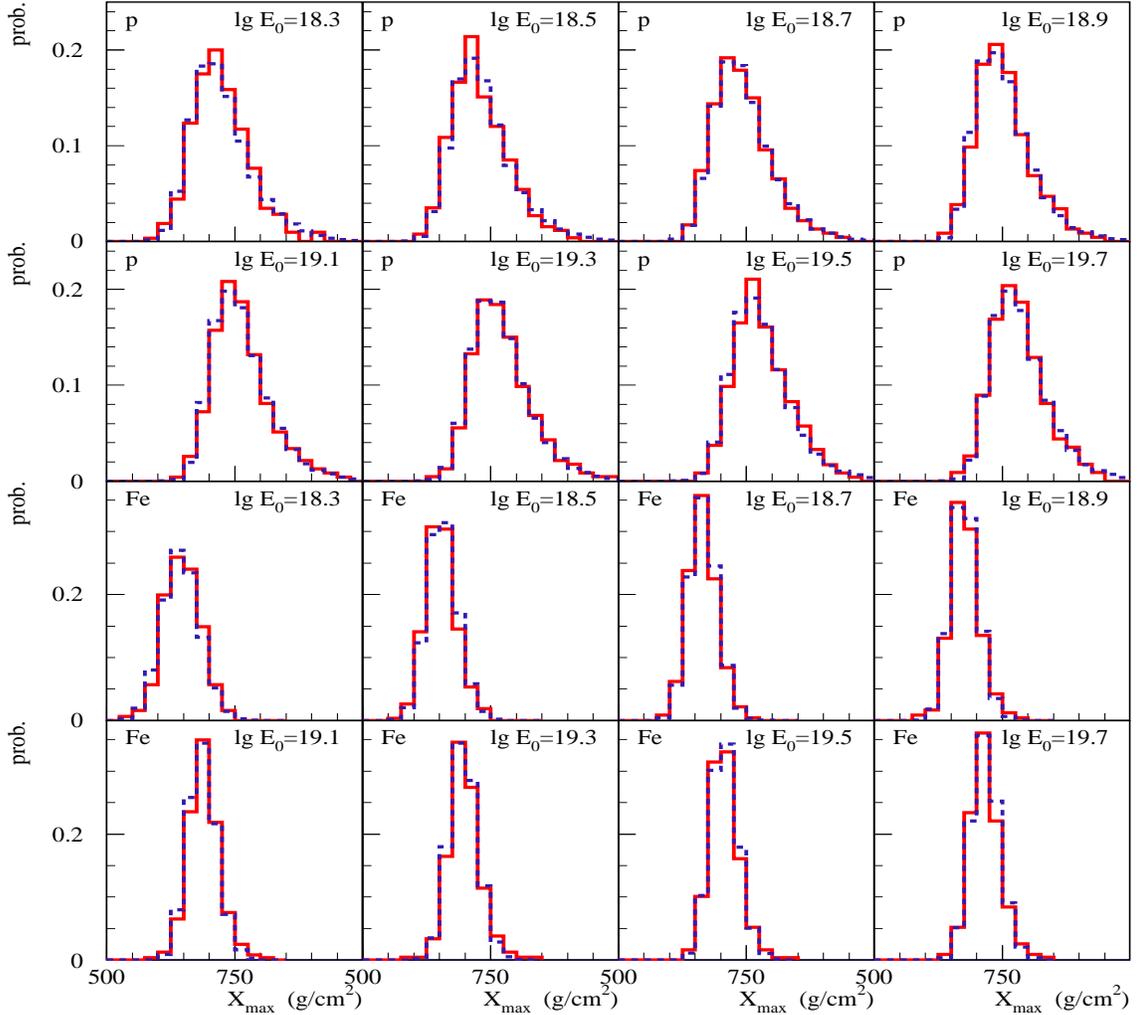}
\caption{$X_{\max} $distributions from TA analysis \cite{ta-xmax} with
 QGSJET-II-03
(full histograms) compared to simulation results with $X_{\max}$-shift
$\Delta X_{\max}$ and Gaussian smearing $\Delta\sigma$ applied (dashed
histograms).\label{fig:xmax-shifted}}
\end{figure*}

In more detail, we applied uniform (energy-independent) shifts
 $\Delta X_{max}^{p}=-25\:{\rm g/cm}^{2}$
and $\Delta X_{max}^{Fe}=-21\:{\rm g/cm}^{2}$ in case of $p$-induced
and $Fe$-induced EAS respectively while adjusting the Gaussian smearing
width $\Delta\sigma$ individually for each primary energy bin, see
Table \ref{tab:sigma}. 
\begin{table*}[tbh]
\begin{tabular*}{1\textwidth}{@{\extracolsep{\fill}}ccccccccc}
\hline 
{\small $\lg E_{0}$} & {\small 18.2-18.4} & {\small 18.4-18.6} & {\small 18.6-18.8} & {\small 18.8-19.0} & {\small 19.0-19.2} & {\small 19.2-19.4} & {\small 19.4-19.6} & {\small 19.6-19.8}\tabularnewline
\hline 
\hline 
{\small $\Delta\sigma_{p}$} & {\small 21} & {\small 20} & {\small 19} & {\small
12} & {\small 14} & {\small 18} & {\small 19} & {\small 13}\tabularnewline
{\small $\Delta\sigma_{Fe}$} & {\small 28} & {\small 19} & {\small 18} & {\small
17} & {\small 19} & {\small 18} & {\small 17} & {\small 18}\tabularnewline
\hline 
\end{tabular*}\caption{Applied Gaussian smearing width $\Delta\sigma$ (in g/cm$^{2}$) for
different primary energy bins for $p-$ and $Fe-$induced EAS.\label{tab:sigma}}
\end{table*}
$X_{\max}$ distributions obtained this way (using QGSJET-II-03) are
compared to TA simulation results in Fig.~\ref{fig:xmax-shifted}.

\end{document}